# PHASE CHARACTERISTICS OF REFLECTING AND TRANSMITTING TYPE TWISTED NEMATIC SPATIAL LIGHT MODULATORS

M B Roopashree[1], Akondi Vyas[1,2], Ravinder Banyal[1] and B R Prasad[1]
[1]Indian Institute of Astrophysics, 2nd Block, Koramangala, Bangalore
[2]Indian Institute of Science, Bangalore
roopashree@iiap.res.in, vyas@iiap.res.in, banyal@iiap.res.in, brp@iiap.res.in

**Abstract:** The phase characteristics of reflecting and transmitting type twisted nematic liquid crystal based Spatial Light Modulators (SLMs) were measured using interferometry. Device parameters like contrast, brightness, input and output polarizer angles have been optimized and SLM phase nonlinearity was reduced by higher order polynomial interpolation. Higher order aberration production ability of SLMs was tested by measuring the shift in the spots of a Shack Hartmann Sensor.

## 1. INTRODUCTION

Spatial Light Modulator (SLM) is a versatile device for reliable and effortless modulation of amplitude and phase of light [1]. They can be used in applications requiring controlled production of phase like phase shifting [2], digital holography [3] and adaptive optics [4]. An accurate calibration of the device is essential before its usage in any controlled phase production application. The phase response of a liquid crystal based SLM is nonlinear. The nonlinearity can be modeled and appropriate command values can be assigned to generate the desirable phase.

A Twyman-Green interferometer was used for the phase measurement in the reflective type SLM case and a Mach-Zehnder interferometer for the transmitting type SLM. The phase characteristics of reflective and transmitting type SLMs were measured at different wavelengths. The phase to gray scale relation depends on the display properties of the Liquid Crystal Display (LCD), namely contrast and brightness. Optimum set of these parameters which allowed the usage of a large grayscale range and gave relatively high amplitude of phase were selected. The phase response at these optimum parameters was then fitted with cubic and tenth degree polynomial interpolation. The nonlinearity of the SLMs was taken into account by using the inverse mathematical expression for the corresponding interpolation polynomials. This linearization procedure of the SLM was checked by addressing grayscale values corresponding to linearly varying phase. This characterization is useful for a controlled and accurate phase production. We checked the production of phase of the reflective type SLM with the shift in the spots of the Shack Hartmann sensor. The variation of the fringe contrast was also measured.

In the second section, the methodology used to measure the phase response of SLMs is explained. Phase measurement results are discussed in detail in the third section. In the last section conclusions are presented.

## 2. METHODOLOGY

The schematic of the Twymann-Green interferometer setup for phase measurement of the reflective type SLM is shown in Fig. 1. The SLM used is LC-R 720 from HOLOEYE. G.P1, G.P2 are two glan polarizer, S.F is a spatial filter setup consisting 40x beam expander and 5μm pinhole, L is 25cm focal length doublet lens for collimation purpose. B.S is a beam splitter, M is a plane mirror in one of the arms of the interferometer, SLM is placed in the other arm. A pulnix CCD camera is used for recording the interferograms. Mellis Griot He-Ne lasers of different wavelengths were used as sources of light.

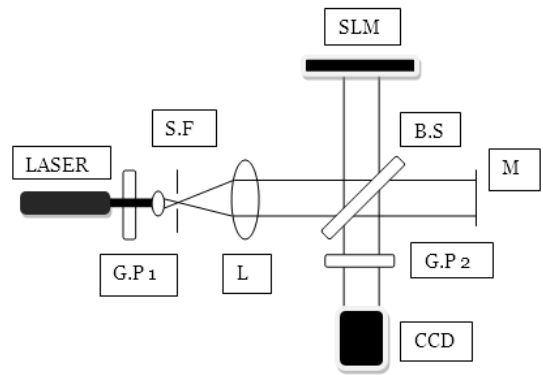

Fig. 1. Twymann-Green Interferometric setup

The schematic of the Mach Zehnder interferometer setup for phase measurement of the transmitting type SLM is shown in Fig. 2. The SLM used is LC 2002 from HOLOEYE. G.P1, G.P2 are two glan polarizers, S.F is a spatial filter setup, B.S1, B.S2 are beam splitters, M1, M2 are plane mirrors, L is a collimating triplet lens with 12.5cm focal length.

Fringe stability is a major problem in the measurement of small phase differences using interferometric arrangement. Vibration isolation table was used for the experiment. Wobbling of the interferograms can occur due to local refractive index fluctuations caused by air. To overcome the wobbling of the interferograms, vertically divided screens on the SLM were addressed as shown in the Fig. 3a. The bottom part of the screen was left dark (0 grayscale) with varying grayscale on the upper part of the screen from 0 to 255 in steps of 8. The resultant interferogram captured on the CCD is shown in Fig. 3b. The interferograms so obtained were smoothened using different image processing techniques. Smoothening was performed by applying medfilt2 and wiener2 filters available in MATLAB. Here medfilt2 stands for 2D median filtering. It reduces salt and pepper noise. This is effective in this case because it simultaneously reduces noise and preserves edges. Another filter wiener2 stands for 2D wiener filter. This is a low pass-filter used to remove constant power additive noise in grayscale images. After smoothening, the measurement of fringe width and fringe shift is straightforward.

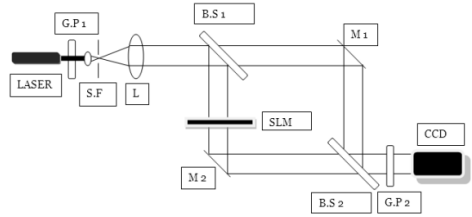

Fig. 2. Mach Zehnder Interferometer setup

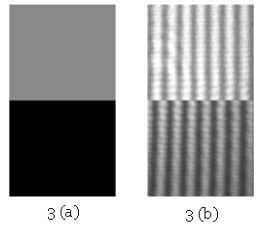

Fig. 3. Vertically divided Screen and the corresponding interferogram

By measuring the fringe width and the fringe shift the amplitude of phase introduced by the SLMs can be calculated using the following formulae,

$$\text{path difference} = \frac{\lambda}{\omega}\delta \qquad (1)$$

where, $\lambda$ = wavelength, $\omega$ = fringe width

and $\delta$ = fringe shift

$$\text{phase difference} = \frac{2\pi}{\lambda} \text{path difference} \qquad (2)$$

### 3. PHASE MEASUREMENT RESULTS

The maximum phase of a liquid crystal based SLM depends on the refractive index of the liquid crystal material, the thickness of the liquid crystal and the wavelength of the source used. Since there is no control over liquid crystal thickness, the phase to grayscale relation can be measured at different wavelengths. The phase response of the reflecting and transmitting type SLMs at different wavelength is shown in Figs. 4 and 5.

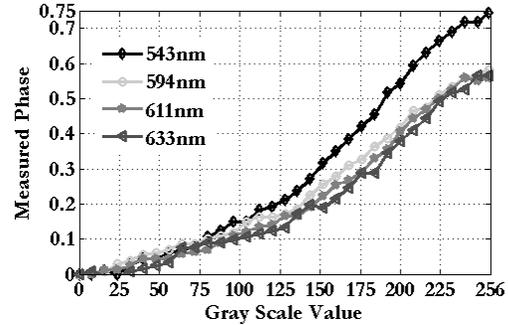

Fig. 4. Wavelength dependent phase response of LC-R 720 SLM

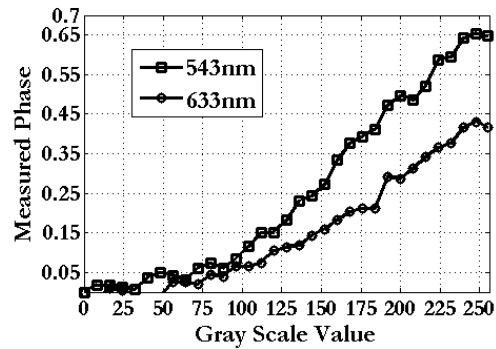

Fig. 5. Wavelength dependent phase response of LC 2002 SLM

The phase response varies with the applied contrast on the SLM liquid crystal screen. Contrast ratio is defined as the ratio of the maximum intensity to minimum intensity. The manufacturer provides a contrast control that can be varied from c=0-255 for transmitting type SLM and c=0-100 for reflecting type SLM. The results of changing phase response with contrast are shown in Figs. 6 and 7. In all graphs, phase is always expressed in wavelength units. The wavelength is specified in corresponding figure caption.

In the SLMs, an allowance was made to adjust the LCD bias voltage. This adjustment controls the contrast ratio of the display device, and this voltage needs to be optimized for best amplitude and phase modulation. Higher contrasts which need development of large voltage difference for small grayscale change leads to saturation effects. On the other hand, low contrasts fail to produce appreciable phase differences. The measured optimum contrast for both of the SLMs lies in the center of the contrast range. It was observed that the brightness change of the SLMs merely allows amplitude modulation and has minimal effect on phase modulation. The input polarization angle for both the SLMs was chosen to be $P=135^0$. For the transmitting type SLM, an analyzer $A=90^0$ was used for best performance.

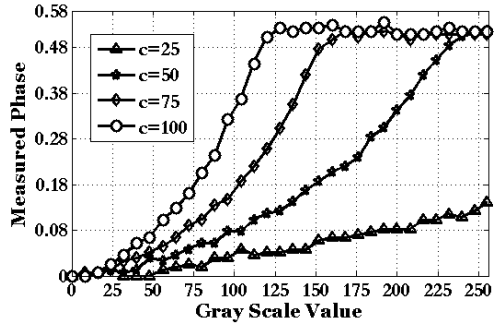

Fig. 6. Contrast dependent phase response of LC-R 720 SLM @ 633nm

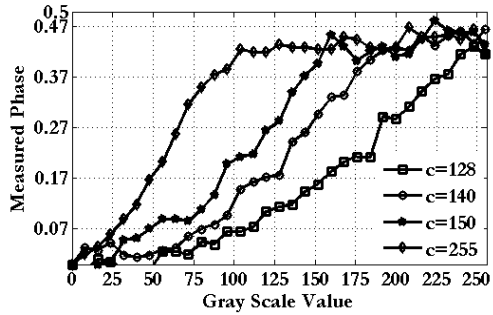

Fig. 7. Contrast dependent phase response of LC 2002 SLM @ 633nm

The phase response of the transmitting type SLM at 543nm was fitted using polynomial interpolation. The resultant equations for cubic and $10^{th}$ degree polynomial interpolation are shown in equations (3) and (4).

Cubic:

$$g = 1713.82p^3 - 1989.02p^2 + 933.18p + 15.82 \quad (3)$$

$10^{th}$ degree:

$g = 24567448.7852107p^{10} - 90450261.0235242p^9 +$

$14315200.63036099p^8 - 126905631.7662587p^7 +$

$68820318.1273422p^6 - 23391774.2662256p^5 +$

$4892722.419124p^4 - 584553.6338582p^3 +$

$31322.9244018p^2 + 495.2425436p + 9.1950276 \quad (4)$

Here 'g' stands for the grayscale value and 'p' stands for the phase value in wavelength unit.

Corresponding to the phase magnitude, $(0-0.65)\lambda$ in steps of $0.05\lambda$, we computed the grayscales using the above formulae and experimentally measured the phase as a re-check for the polynomial interpolation. The corresponding linearized plots for cubic and $10^{th}$ degree interpolation are shown in Figs. 8 and 9.

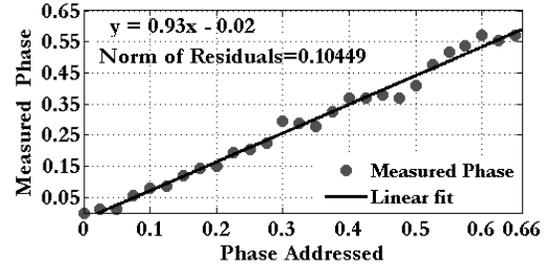

Fig. 8. Cubic inversion check for transmitting type SLM @ 543nm

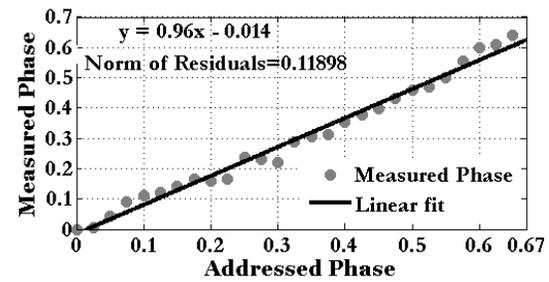

Fig. 9. $10^{th}$ degree polynomial inversion check for transmitting type SLM @ 543nm

The phase response of reflecting type SLM at 633nm was fitted using polynomial interpolation. The resultant equations for cubic and $6^{th}$ degree polynomial interpolation are shown in equations (5) and (6).

Cubic:

$g = 2315.3876p^3 - 2469.6140p^2 - 1096.8648p$

$\qquad + 14.2464 \quad (5)$

$6^{th}$ degree:

$g = -57059.6360p^6 + 132609.9681p^5 - 110176.3732p^4$

$\qquad - 43431.1814p^3 - 9326.8085p^2$

$\qquad + 1516.2394p + 10.0568 \quad (6)$

The linearization results are plotted for cubic and $6^{th}$ degree interpolation and shown in Figs. 10 and 11.

The phase to grayscale relation was verified using Diffractive Optical Lens (DOL) based Shack-Hartmann Sensor (SHS) realized using SLM. Linear tilts of increasing magnitude were applied across sub-apertures of SHS and the shift in the spots was measured. The linear relation shown in Fig. 12 between the tilt and the shift in the spot confirms a proper phase characterization of the SLM.

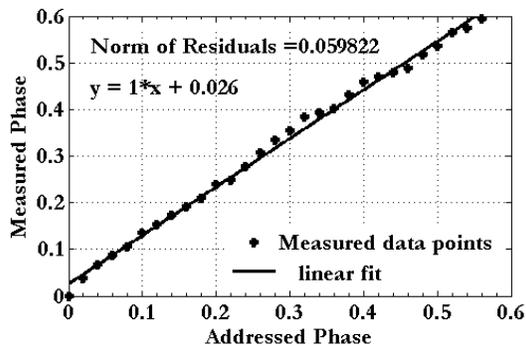

Fig. 10. Cubic inversion check for reflecting type SLM @ 633nm

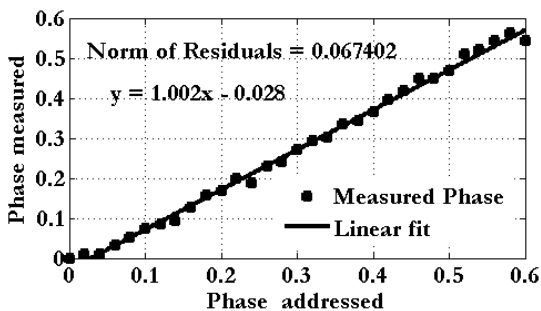

Fig. 11. 6[th] degree inversion check for reflecting type SLM @ 633nm

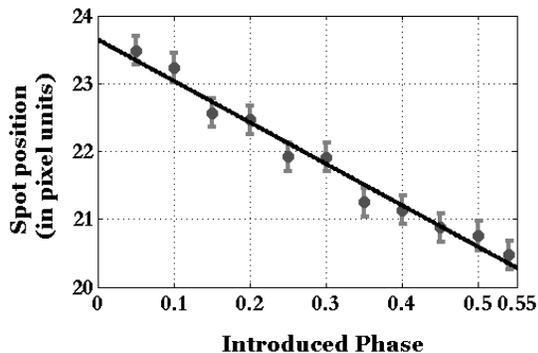

Fig. 12. Relation between applied phase and shift in the spot of SHS @ 633nm

### 4. CONCLUSIONS

Changing brightness of the SLMs had minimal effect on the phase modulation characteristics. Very high and too low contrasts either led to saturation effects or production of low phase. The optimum contrast for both SLMs is at the center of the contrast range. This observation can be attributed to the voltage difference dependent contrast of the display. Hence rest of the analysis was performed using 50% contrast. LC-R 720 was found to give a maximum phase of 4.52 ± 0.01 radians at 543nm and 3.58 ± 0.01 radians at 633nm. The maximum phase for LC 2002 was measured to be 4.15 ± 0.01 radians at 543nm and 2.76 ± 0.01 radians at 633nm. The input polarizer was fixed at $45^0$ which is the orientation of molecular director of the nematic LC-SLMs. At output polarizer angles of $0^0$ and $45^0$, the magnitude of phase was significant for LC-R 720. In the case of LC 2002, the optimum analyzer angle was found to be $90^0$.

The obtained nonlinear phase curves were fitted using cubic interpolation. Inverse transformation was performed to obtain expressions for grayscale as a function of applied phase. Cubic inversion has a linearization residual error of ±0.19 radians in the case of LC-R 720 and ±0.31 radians for LC 2002. It was observed that inversion using higher order interpolation reduces the residual error. The measured shift in the spots of SHS corresponding to the applied phase difference was found to be linear within the experimental errors ascertaining the possibility of using SLM for higher order aberration production and compensation in adaptive optics testing.